# AI-assisted 3D Preservation and Reconstruction of Temple Arts

Naai-Jung Shih

Department of Architecture

National Taiwan University of Science and Technology

43, Section 4, Keelung Road, Taipei 106, Taiwan

shihnj@mail.ntust.edu.tw

**Abstract:** How does AI connect to the past in conservation? What can 17 years old photos be helpful in a renewed effort of preservation? This research aims to use AI to connect both in a seamless 3D reconstruction of heritage from imagery data taken from Gongfan Palace, Yunlin Taiwan. AI-assisted 3D modeling was used to reconstruct correspondent details across different 3D platforms of 3DGS or NeRF models generated by Postshot or KIRI Engine. Polygon or point models by Zephyr were referred to and assessed in two sets. The results also include AI-assist modeling outcomes in Stable Diffusion and Postshot-based animation. The evolved documentation and interpretation in AI presents a novel arrangement of working processes contributed by new structure and management of resources, formats, and interfaces, as a continuous preservation effort.

## 1. Introduction

Gongfan Temple, or Gongfan Palace, is located in Mailiao Township, Yunlin County, Taiwan. The name refers to the peace and prosperity of all living beings in the area. It has worshiped the Six Mazus since 1685. For a history of 340 years, it is the earliest temple in Taiwan to worship Mazu. In 2006, the Yunlin County Government announced it as a county-designated historic site, in memory of the tangible and intangible heritage. The preservation of cultural assets contributes to the policy jointly advocated by advanced technologies for the exquisite carvings installed interiors and exteriors (see Figure 1). How to effectively preserve assets and collaborate with old resources is one of the current priorities of cultural policies.

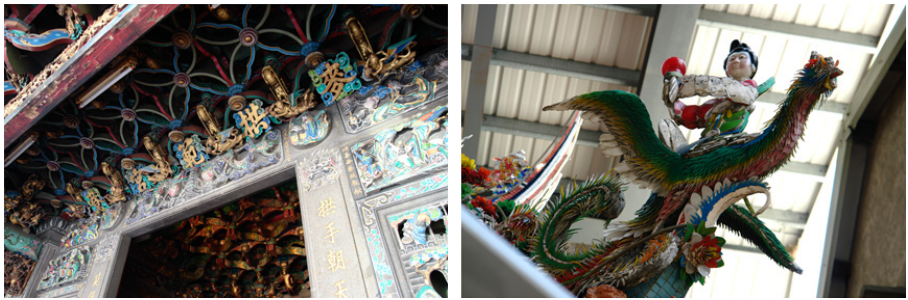

**Figure 1.** Gongfan Palace and decorations.

Temple is frequently decorated by 3D creations inspired by ancient stories for educational purposes. The composition of traditional stories is often designed by craftsmen in delicate pro-tagonists, expressions, body shapes, and scenes in 3D or 2.5D. People can further distinguish the 3D narratives or identities of different craftsmen from the tension of the body or the wrinkles on clothes. The narratives of 3D scenes vitalize old memories nowadays.

This research aims to connect AI and conservation in seamless 3D reconstruction of heritage from imagery data taken 17 years ago. The as-built physical and digital data are mutual influential in preserving the sustainability of cultural heritage. The temple was digitally preserved using 3D laser scanning technology in 2008 [1], associated with various working photos recorded earlier. How can the 17 years old photos be helpful to renew the effort of preservation? Since AI has changed how data can be sustainably generated now, how it connects to the past in a different paradigm shift remains to be seen in conservation.

Upon the new development of photogrammetry and AI, the efficiency of 3D reconstruction has been improved. The images should be reused to reconstruct the temple and decorative arts, such as wood carvings, stone carvings, clay sculptures, cut-outs, murals, tablets, plaques and other building components at that time.

*1.1. Related studies*

Survey shows 3D Gaussian Splatting (3DGS), which represents a paradigm shift in neural rendering, has potential to become a mainstream method for 3D representations. It effectively transforms multi-view images into 3D Gaussian in real-time [2]. 3DGS produces highly detailed 3D reconstructions. Software can estimate camera poses for arbitrarily long video sequences [3]. In surface reconstruction of large-scale scenes captured by UAV, the quality of surface reconstruction was ensured in heavy computational costs [4]. The high-precision real-time rendering reconstruction was also applied to cultural relics and buildings in large scenes [5], peach orchard [6], or historic architecture using 360° capture [7]. In addition to be part of the massive 3D digitization of the remains after the Notre-Dame de Paris' fire [8], neural rendering was applied to leaf structure [9], the analysis and promotion of dance heritage [10], or as an ethical framework for cultural heritage and creative industries [11]. Special renderers were developed for effective visualization of point cloud or mesh [12]. RODIN applies a generative model for sculpting 3D digital avatars using diffusion [13], with 3D NeRF model presents computational efficiency.

3D content generation has benefited by advanced neural representations and generative models [14]. For a relatively small scaled object or cavity, a standard smartphone can be used to capture tangible cultural heritage using only RGB images in affordable and easy deployment [15]. The accuracy of the scaled 3D models created with the iPhone and commercial structure-from-motion (SfM) applications met the inspection requirements of crane hook [16]. For details, generative AI was used to create accessories of virtual costume [17].

When the detail and accuracy has to meet medical requirements, photogrammetry was conducted for 3D medical didactics, based on mobile phone apps which are accessible, user-friendly, and potentially for VR and AR [18-20]. Its application provides comparable scanning capabilities at a significantly low cost for medical applications like intraoral scan [21]. KIRI Engine was used for image capture in designing a patient specific helmet [22] and acquiring 3D meshes of cleft palate models [23]. KIRI program contains more accurate details, faster, and does not require computer specifications or time-consuming processing [24]. Although the accuracy of models could be lower, smartphone accessibility and practicality proved to be significant advantages [25].

If smartphone scan applications can be applied for intraoral scan and costume details, it should also be feasible for documentation of cultural heritage in architecture. UAV with 1" sensor of camera can document images to reconstruct the scale of a large site. A smartphone with reso-lution of 200 million pixels should be feasible for detailed reconstruction of cultural heritage in small scale.

## 2. Materials and Methods

The continuous update of heritage data is the most direct solicitation to conservation conducted 17 years ago. This preservation consists of a series of 3D reconstruction processes using scans, photogrammetry, or AI (Figure 2). AI-assisted software and cloud computing are applied (1.e. KIRI Engine). For desktop computing, Postshot creates model to solve the issue of data privacy and sovereignty, in a basic setup of Nvidia GPU RTX 2060. The working platform changes and produces a 3D model that is different from the traditional one. It opens up the diversity of subsequent operating platforms to game engine (Unreal Engine or UE) and multimedia software (After Effect or AE). The original restrictions in taking pictures determine the selection of subsequent applications, such as Postshot, KIRI Engine (at least 20 images, software specified) [25], and Zephyr (Figure 3). Progress has been made in 3D reconstruction with the combination of AI. 3DGS has substantially improved in reality and output convenience.

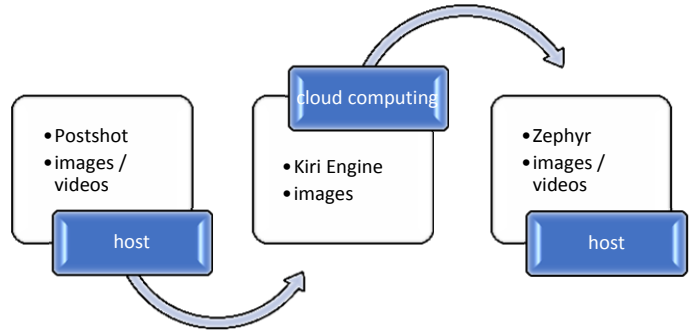

**Figure 2.** 3D reconstruction process and the preferred number of images.

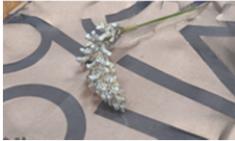
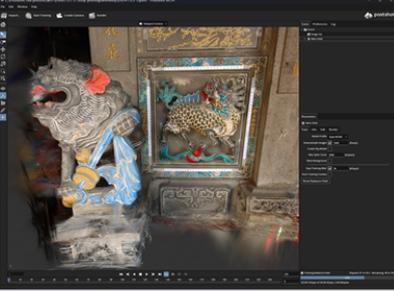
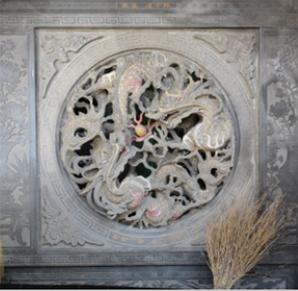
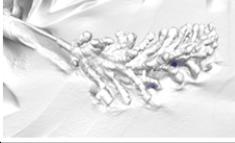

| | | | |
|---|---|---|---|
| **purpose** | detail test | element test | element test |
| **data** | 3DGS mesh | 3DGS | mesh |
| **software** | KIRI Engine | Postshot | Zephyr |

**Figure 3.** Tests for different purposes, data types, and applications.

AI-assisted 3D model reconstruction verifies correspondent details across different 3D platforms of solutions. Point cloud and mesh by Zephyr were referred to and compared with 3DGS or polygon models generated by Postshot and KIRI Engine under the following requirements.

1. Availability of sufficient imagery data: Reconstruction may fail for limited number or chaotic arrangement of image sequence.

2. Comparison of cross-program reconstruction results in the same level of details: Thin element must present roughly the same level of details, and should not disappear, mis-create, or merge with adjacent parts.

3. Image segmentation: Unclear image may generate unexpected 3D results. For example, the dark horns of a cow were not generated because of its similarity to dark background color. In contrast, an extruded nose was generated out of similar background color.

4. Support the options for generating polygon models: Polygon or mesh model was still requested additionally for following-up study or visualization using shaded/pbr (physical based rendering), MTL, and AR, based on OBJ or 3DGS (with/out mesh) option.

5. 3D program accessibility: Variation exists in desktop/mobile configuration or subscription plan by polygon numbers, original image size, and texture resolution.

6. Seed implementation: Seed value was available to record training and extend generation experiences.

The limitations of this study include the completeness of old images. About 1,500 old photos (Figure 4) were maintained and used under the categories of working records and environment observation, taken in radial and centripetal shots. Reconstruction restrictions include the number of photos, how the photos were taken, and the quality of the photos. A large number of pictures included striped survey markers placed temporarily next to the target for alignments in registration. The markers caused interference to 3D reconstruction. Fortunately majority numbers of the markers were removed either automatically or manually.

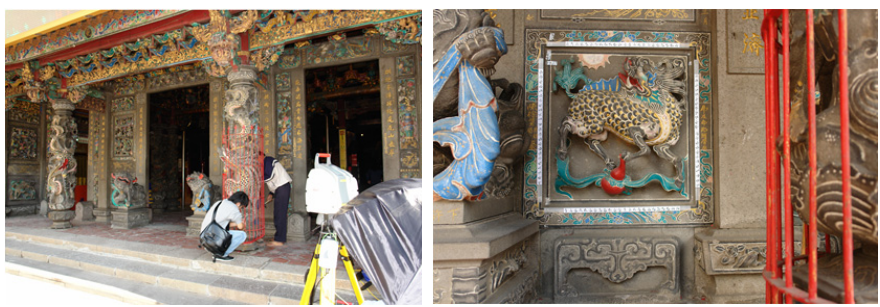

**Figure 4.** Original images.

In total, about 1500 images were taken using old SONY DSC-R1 in 3888x2592 pixels, 24 bit, and 72 dpi resolution, with some taken using Panasonic DMC-FX100 in 4000x3000 pixels, 24 bit, and 72 dpi resolution. The success of reconstruction was determined by the proportion of available photos that generated 3D models or the number of failures. The 3D output rate was about 80%, and the satisfactory success rate was about 50%, but the level of reality was still quite helpful from a sustainable perspective of digital data.

## 2.1. Related studies

Initial period of preservation was conducted between 2008.8 and 2009.3, implementing Leica HDS 3000 laser scanner and Cyclone software for long-range scan, Optix 400 RealScan USB and Color_Calibration software for detailed scan (Figure 5). The former completed a total of 612 scans and 74,584,968 points from 33 scanning locations on ground levels and commanding heights. The latter completed 29 components and 1023 scans. For wood carving, 34 pieces, 154 Million polygons, 1029 scans were made in former attempt.

The main temple was under renovation and covered under a weather canopy, causing partial shielding for long-range scan. Near-range scan data must be calibrated for the first use, and subjected to limited space accessibility and interference from light sources. After completing the field scan work, different scans need to be registered, sampled, noise filtered, and color corrected. There were often small areas that could not be recorded.

The temple has numerous fine components under the same category with similar configuration. To avoid mis-identification, a customized coordinate system was deployed for vertical, horizontal, and elevation positioning, based on existing partitioning modules. The final results were integrated with Metadata definition in web pages (Figure 6).

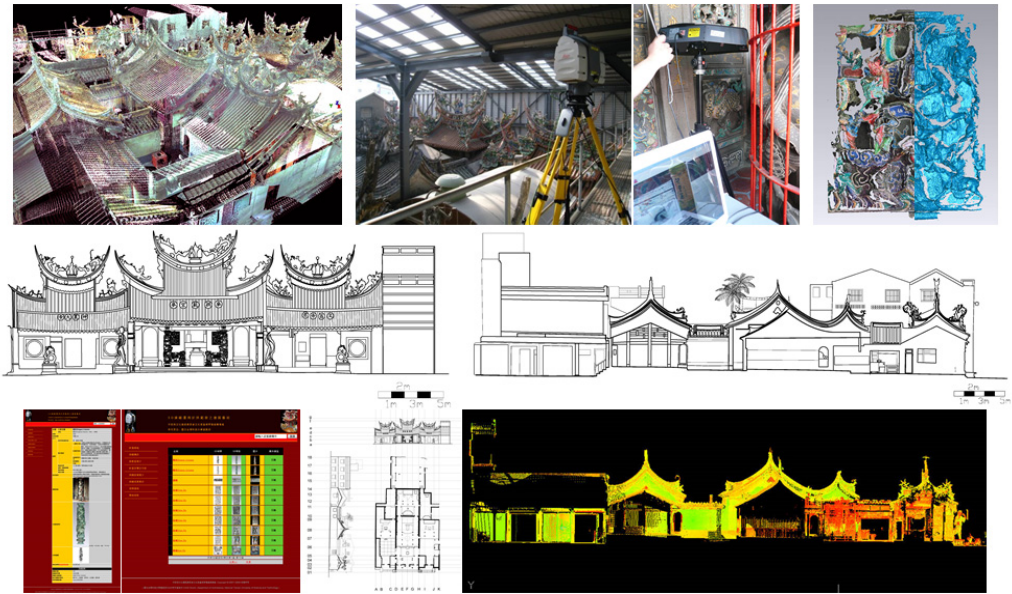

**Figure 5**. Leica HDS 3000 scanner and 3D point cloud model, Optix 400 RealScan USB and 3D mesh model with/out texture (top); traced vector drawings of elevation and section (middle); web page of Metadata (dragon pillar) and element list, customized coordinate system, section made from point cloud (bottom)

## 3. Results

3D models were reconstructed using Postshot, KIRI Engine, or Zephyr. The results were compared, assessed, or animated.

### 3.1. AI-assist modeling in 5 types

The results are categorized into 5 types of components (Figure 6).

1. Stone carving: The main components of stone carving were made to dragon pillars, stone lions, and decorations on walls, including locations on top (top block), body (torso block), and skirt (apron block). The carving methods include openwork, relief, and round carving. Postshot and Zephyr were used mainly, while specific components were modeled and compared using all the applications.

2. Temple structures: This category included 5 models made to Sanchuan Hall, roof, and interior. The old photos recorded during the 3D scanning process were taken before the renovations were complete on the corners at front right and left rear. Zephyr and Postshot were used to reconstruct the site using image set(s) sequentially rearranged in a new order out of original shooting angles. In the temple front, the red iron fence was removed from the dragon pillar. Although it was obstruction-free, the model resolution was lower than expected, since the images were taken in a distance and less than 5 images were available.

3. Wood carving: The wood carving included brackets, queti, drop beam, and hanging canister. Selection was based on the iconic representative of the character introduced on the old official web pages. Zephyr was mainly used to reconstruct the models.

4. Clay sculpture: Most of the clay components are located on the wall below the roof ridge. In order to avoid damaging the building components or causing danger to the workers during the operation, components used to be selected for safer operation. Limited accessibility to the back view prevented a thorough scan of the subject.

5. Cut-and-paste craft: Colored bowls are processed into pieces and then pasted to form 3D shapes.

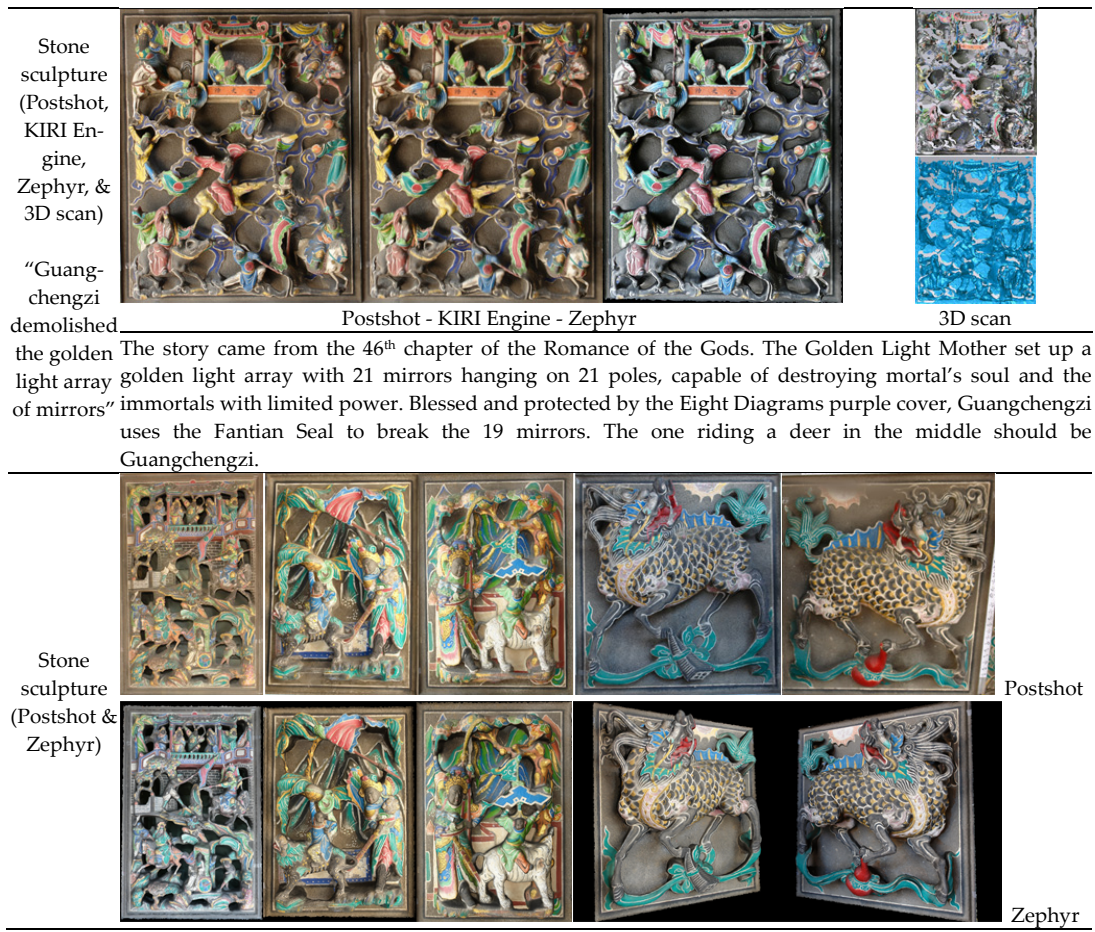

| Stone sculpture (Postshot, KIRI Engine, Zephyr, & 3D scan) "Guangchengzi demolished the golden light array of mirrors" | Postshot - KIRI Engine - Zephyr | 3D scan |

The story came from the 46th chapter of the Romance of the Gods. The Golden Light Mother set up a golden light array with 21 mirrors hanging on 21 poles, capable of destroying mortal's soul and the immortals with limited power. Blessed and protected by the Eight Diagrams purple cover, Guangchengzi uses the Fantian Seal to break the 19 mirrors. The one riding a deer in the middle should be Guangchengzi.

| Stone sculpture (Postshot & Zephyr) | | Postshot |
| | | Zephyr |

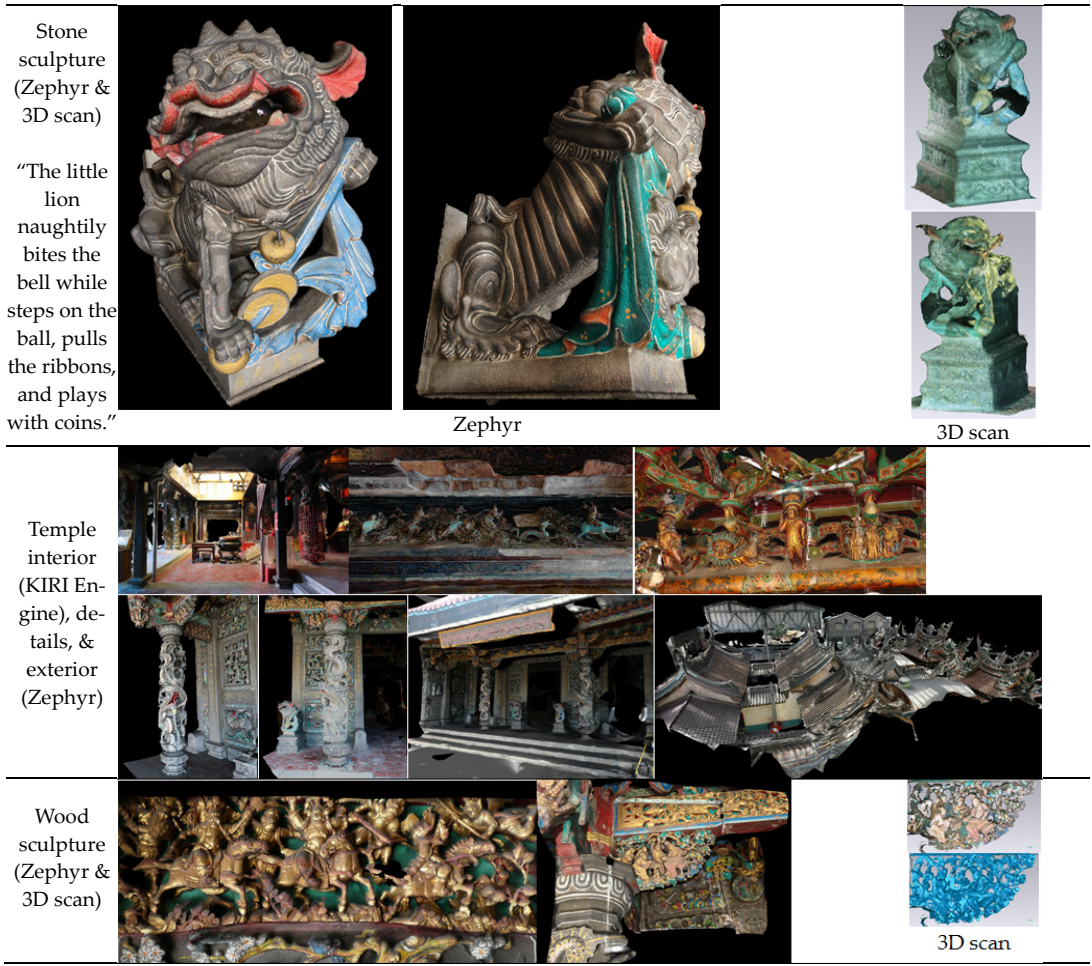

**Figure 6**. The 5 types of components and models.

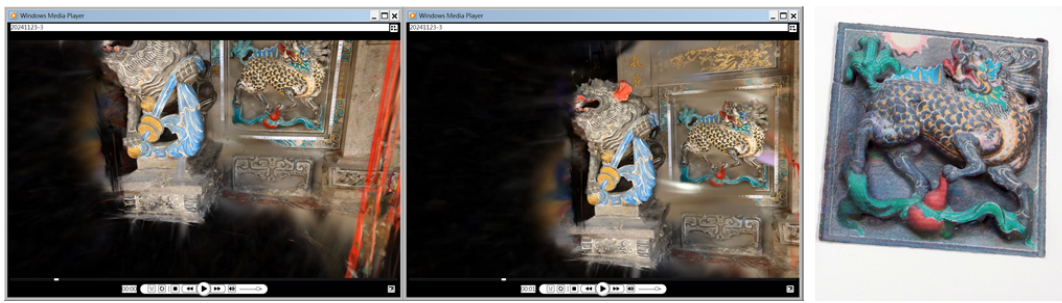

**Figure 7.** Postshot animations (left) and 3D prints (right).

### 3.2. Postshot animation and 3D prints

AI-assisted modeling, which was conducted using off-shelf ready-made applications, also supported animation in Postshot and Smartphone AR in Augment. The 3D model of the Gongfan Palace was scanned using reverse engineering technology in point model. In order to maintain details, the reduction of polygon numbers was usually a trade-off between the computing speed and the best details to be presented. Postshot directly works from images to animations (Figure 7, left) regardless of polygon counts. It creates realistic scenes in basic hardware setup, and supports game engines (Unreal Engine) and multimedia time track (After Effect).

3D print is an effective physical representation of data. ComeTrue T10, 1200 x 556 dpi), which replaces Prodigy Plus, was used to print 3D color models. Inkjet dyes were printed on layers of gypsum-like powder to verify texture and the relations of geometric attributes between parts (Figure 7-right). The detail presented in the old reconstruction was improved to a fine appearance by increasing the model resolution for self-explanatory visual and structural details.

## 4. Discussion

Preservation needs constant maintenance, since the demand, format, nature, interference would be different or evolved in the future. Different approaches not only just created novel types of data, but enabled opportunity with new interfaces and platforms. 3D contents, which used to be a collative presentation of heterogeneous types of data, is enriched by AI-assisted generation.

### 4.1. 3D assessments

3D models were cross-assessed between the outcomes made by KIRI Engine, Postshot, and Zephyr. Larger tolerance was encountered in the following three specific cases, although all the models presented certain level of fidelity. The standard deviation was less in KIRI Engine-Zephyr paired set (Figure 8, top), than that made by Postshot-Zephyr paired assessment (Figure 8, bottom). Instead of 3DGS, the PLY format was exported from Postshot, converted into 3D polygons in Geomagic Studio, and globally registered with the model created by Zephyr. The PLY model was in unevenly distributed splatters.

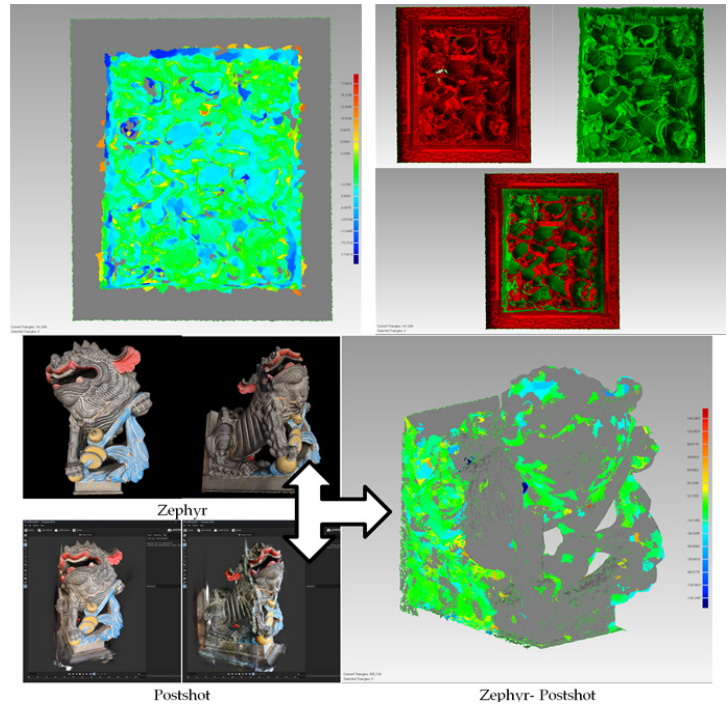

**Figure 8.** Two types of assessment were conducted: KIRI Engine vs. Zephyr (top: Max. Distance: 17.80 mm, Average Distance: 3.23 mm, Standard Deviation: 2.93 mm), Zephyr vs. Postshot (bottom: Max. Distance: 140.25 mm, Average Distance: 19.13 mm, Standard Deviation: 21.69 mm.

The second comparisons were made by registering two overlapping 3D models and assessed in terms of standard deviation, maximum distance, or average distance. Global registrations were made within the trimmed boundary of the focused subject, following manual registrations conducted to the two models in original size.

### 4.2. 3D composition analysis by sections

In order to verify the level of openwork, sections parallel to the wall surface were made to inspect if thin elements and wall background were really detached. This is one of the distinguished stone sculptures which created flows and tensions between interlaced layers of characters and subjects within a short depth (Figure 9). The two test sets were reconstructed by KIRI Engine and Zephyr, in which the latter presented more details than the former. The depth-based slices, which were made in Cloudcompare, deconstructed the sculpture composition by layers. The composition was an interlaced complex of the name badge in the center-top, the main fighting character cluster on the center and lower left, the counter balanced character on the center-right fighting with two mirrors in hands, and the clouds. The composition presents tension between character body features, riding beasts, weapons, flags, and clouds. The cloud had three hierarchical layers from paint on background, elevated curves from background, to extrude with the same layer of characters. The most elevated arrangement part was located on the left-top, with most parts of the characters merged into the background and behind clouds.

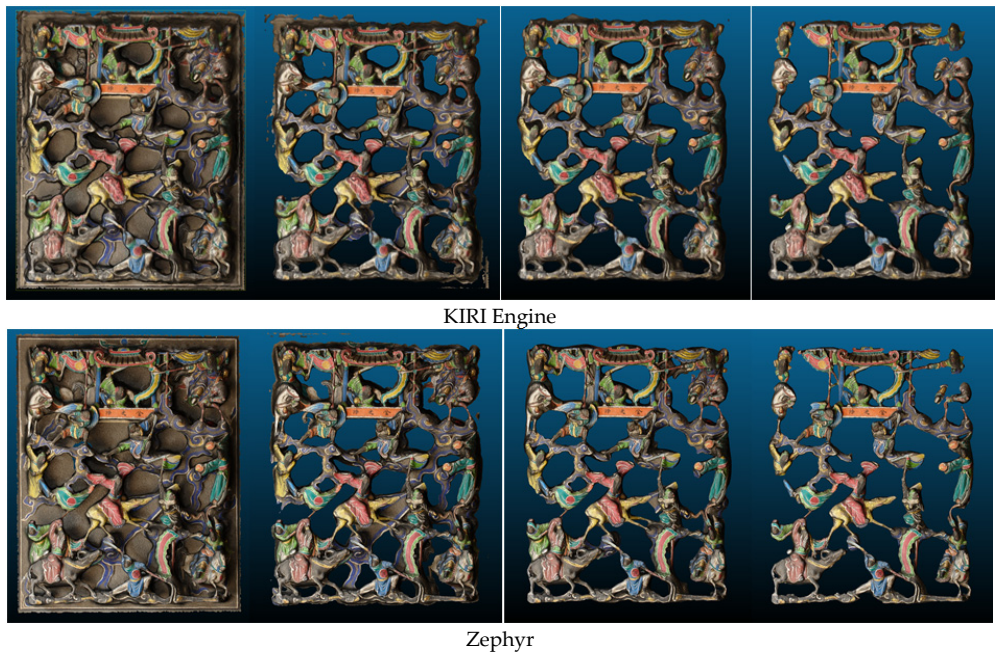

**Figure 9.** Analysis of flows and tensions between interlaced l of characters and subjects by sections within a short depth in KIRI Engine (top) and Zephyr.

### 4.3. New structure and management of resources, formats, and interfaces

A well-documented set of images can be used as reference to support collaborative applications. The level of details was restricted by the available angles provided by old images. The tra-dition photogrammetry approach, i.e. Zephyr, not only generates a more favored appearance. In other words, all 3D data were mutually referenced.

Although the afterward creation regards old imagery data, AI-based assistance contributed to a new structure with more feature-inspired management (Table 1). The structure also enlightened a more diversified interface enabling situated application and delivery, for example, 3D print and AR.

The adaptive reuse of old photos in AI associated with new interpretation in fidelity. Compared to the large and heavy physical setting of a 3D scan system, the image-based modeling is an efficient reconstruction alternative. In contrast to traditional VR and animation, polygon-free manipulation was made possible directly in gaming engines or multimedia platforms. The management of 3D data was facilitated by a 3D cloud platform for AR interaction.

**Table 1.** Comparison of old and new approaches

| | Old | New |
|---|---|---|
| Management | 3D scans, point cloud models | 1500 images |
| Platforms of software | Cyclone, Geomagic Studio | Postshot, KIRI Engine |
| Old tools | • Leica HDS 3000 laser scanner: 360° x 270° field-of-view, optimal effective range 1m-200m, sngle point accuracy surface precision 6mm<br>• Optix 400 RealScan USB: precision per point: ±50 microns @ 200 mm, depth of field: 400-600 mm, 30° field of view | • Sony DSC-R1 in 3888x2592 pixels, 24 bit, 72 dpi<br>• Panasonic DMC-FX100 in 4000x3000 pixels, 24 bit, 72 dpi |
| 3D creation / Generation | LiDAR, time-of-flight (TOF) | • NeRF training and Gaussian Splatting techniques (Postshot)<br>• 3D Gaussian Splatting, Neural surface reconstruction (KIRI Engine)<br>• Photogrammetry (Zephyr) |
| Media / interaction | • Web pages + 3D plugin<br>• FDM 3D print | • Animation in splatters<br>• Color RP (3DP)<br>• AR interaction (Augment)<br>• PBR option |
| Internet delivery | Web pages | • AR model in cloud access<br>• Cloud computing or training (KIRI Engine)<br>• Host computing (Zephyr, Postshot) |
| Format + interface | Pts, obj, dwg, jpg | 3DGS w/out mesh, obj, pbr |
| 3D interaction | VR (Quest) | • AR: Augment<br>• Internet view: Sketchfab<br>• Gaming support: Unreal Engine<br>• Media support: After Effect |
| Output in general | • 34 pieces of artifacts + temple<br>• 1029 scans<br>• 154 Million polygons<br>• Vector drawings<br>• Web pages<br>• Quest VR model<br>• 1500 images | • 3D models in 3DGS w/out mesh, obj, pbr<br>• Images from Stable Diffusion<br>• Color 3D print<br>• Animation from Postshot |
| Physical output | 1 ABS print | 2 3D color prints |
| Effort | • 299 days<br>• 11 people, 443 people-day | |

Evolved technology comes from adaptive reuse and acknowledgements of cultural heritage. In other words, this is a continuous effect which evolves the preserved artifacts with inspired outcomes. The Postshort file currently only works specifically in Unreal Engine, Blender, or After Effect. KIRI Engine, which provides an option of 3DGS with/out mesh, solves the problem. Eventually, the 3DGS can be 3D reformatted not just in PLY, but also for 3D print and AR.

## 5. Conclusions

The old photos are used as the main objects of operation for a sustainable effort and resilient attempt in conservation. Do we use powerful AI just to create general purpose 3D scenes for games or media, by skipping the formal polygon model? The fly-through animation no longer needs a model-creation process! So is the platform to elaborate the dynamic effect a presentation needs. Let alone the details and realistic scenes are far better without polygons than those used to be.

AI-assisted 3D reconstruction process inspires to rethink and verify gray area in character and craftsmanship. No matter if it is an openwork carving or 2D image made by vector drawing, 3D reconstruction has proven feasible even using photos taken 17 years ago.

The reconstruction of openwork carvings and details has created curved thin elements, like cuffs and ribbons around the body, and reinterpreted outcomes for heritage vocabularies inspection. The working process with AI opens up to wider selections and descriptions of format (3DGS, for example) for heritage Metadata.

The future works should be better conducted, considering the restricting of old imagery resources. The continuous effect from Postshot to KIRI Engine has evolved new structure and management of resources, formats, and interfaces of the documentation and interpretation in AI.